\documentclass[english]{article}
\usepackage[T1]{fontenc}
\usepackage{geometry}
\usepackage{babel}
\usepackage{graphicx}

\makeatletter

\newcommand{\LyX}{L\kern-.1667em\lower.25em\hbox{Y}\kern-.125emX\spacefactor1000}

\date{}
\makeatother

\begin{document}
\renewcommand{\figurename}{Fig.}

\title{On Multiphoton Bremsstrahlung}

\author{A. Kolchuzhkin\thanks{
corresponding author. E-mail: amk@chair12.phtd.tpu.edu.ru
}, A. Potylitsyn, A. Bogdanov, I.Tropin}

\maketitle
\centerline{Tomsk Polytechnic University, 634034, Tomsk, Russia}

\begin{abstract}
Kinetic equations for the process of multiphoton bremsstrahlung of electron in a matter 
have been used to derive the equations for the moments of distributions of electrons
over the number of emitted photons and over the energy loss. The equations for
the moments have been solved in continuous slowing down approximation. It has
been shown that the moments of multiphoton distributions can be expressed in
terms of the moments of macroscopic differential cross section of the process.
The results of analytical calculations have been compared with the data of Monte
Carlo simulation.
\end{abstract}

\section{Introduction}

\qquad The bremsstrahlung of electrons passing through a matter is an example of stochastic
process where the number of emitted photons as well as their energies are random.
The importance of multiphoton effects was underlined in \cite{1} where the
problems of experimental study of bremsstrahlung cross section were discussed.
The summary energy of emitted photons, as a rule, is measured in such experiments
and the distribution of electrons over energy loss differs from the energy distribution
of radiated photons \cite{2}. Kinetic equations for the distribution of electrons
over the number of emitted photons and for the distribution over the radiation
energy loss are given in the paper. They are used to get the equations for the
moments of the distributions. Qualitative analysis of the path and energy dependences
of the moments is given in continuous slowing down approximation with simple
relativistic Bethe-Heitler formula for the macroscopic differential cross section
of bremsstrahlung. It is shown that the moments of differential cross section
can be expressed in terms of the moments of energy loss distribution which can
be determined in experimental way. The results of the analysis are compared with
the data of Monte Carlo simulation.

\section{Kinetic equations for multiphoton bremsstrahlung}

\qquad Let us consider an electron of energy \( E \) traveling in a matter the path
length \( l \). The probability \( P(n|E,l) \) to emit \( n \) photons of
bremsstrahlung along the pass \( l \) obeys the balance equation (the Kolmogorov-Chapman
equation) \cite{3}:

\begin{equation}
\label{sec:Pn}
P(n|E,l)=(1-s\Sigma (E))P(n|E,l-s)+s\Sigma (E)\int _{0}^{E}\frac{\Sigma (E_{\gamma };E)}{\Sigma (E)}P(n-1|E-E_{\gamma },l)dE_{\gamma },
\end{equation}
 \( \Sigma (E) \) and \( \Sigma (E_{\gamma };E \)\( ) \) being the total
and differential macroscopic cross sections of bremsstrahlung, \( s \) is a
small part of \( l \) . The first term in the right side of Eq.(\ref{sec:Pn})
corresponds to electrons which pass small path \( s \) without radiation and
\( 1-s\Sigma (E) \) is corresponding probability. These electrons have to emit
\( n \) photons along the rest of the path \( l-s. \) The second term corresponds
to the electrons which emit the first photon passing path \( s \) and \( s\Sigma (E) \)
is corresponding probability. These electrons have to emit \( n-1 \) photons
after that but the energy of electrons equals to \( E-E_{\gamma } \) where
\( E_{\gamma } \) is random energy of the first emitted photon and \( \Sigma (E_{\gamma };E)/\Sigma (E) \)
is the probability density function of \( E_{\gamma } \) . In the limit \( s\rightarrow 0 \)
the formula (\ref{sec:Pn}) gives the integro-differential equation for \( P(n|E,l) \)
\cite{4}:

\begin{equation}
\label{dPn}
\frac{\partial }{\partial l}P(n|E,l)+\Sigma (E)P(n|E,l)-\int ^{E}_{0}\Sigma (E_{\gamma };E)P(n-1|E-E_{\gamma },l)dE_{\gamma }
\end{equation}
 with boundary condition

\[
P(n|E,l=0)=\delta _{no.}\]

In similar way one can obtain the balance equation for the probability density
function \( P(Q|E,l) \) describing the distribution of electrons over the energy
loss \( Q \) :

\[
P(Q|E,l)=(1-s\Sigma (E))P(Q|E,l-s)+s\Sigma (E)\int _{0}^{E}\frac{\Sigma (E_{\gamma };E)}{\Sigma (E)}P(Q-E_{\gamma }|E-E_{\gamma },l)dE_{\gamma },\]
which leads to the equation \cite{4}

\begin{equation}
\label{dPQ}
\frac{\partial }{\partial l}P(Q|E,l)+\Sigma (E)P(Q|E,l)-\int ^{E}_{0}\Sigma (E_{\gamma };E)P(Q-E_{\gamma }|E-E_{\gamma },l)dE_{\gamma }
\end{equation}
with boundary condition

\[
P(Q|E,l=0)=\delta (Q),\]
\( \delta (Q) \) being the Dirac \( \delta  \) -function.

Multiplication of both side of Eq.(\ref{dPQ}) by \( Q^{n} \) and integration
over \( Q \) gives the equations for the moments of the distribution \( P(Q|E,l): \)

\[
\overline{Q^{n}}(E,l)=\int ^{E}_{0}Q^{n}P(Q|E,l)dQ.\]
 The equations for two first moments have the form \cite{4}:

\begin{equation}
\label{sec:Q1}
\frac{\partial }{\partial l}\overline{Q}(E,l)+\Sigma (E)\overline{Q}(E,l)-\int ^{E}_{0}\Sigma (E_{\gamma };E)\overline{Q}(E-E_{\gamma },l)dE_{\gamma }=\beta (E),
\end{equation}
 
\[
\overline{Q}(E,l=0)=0,\]

\begin{equation}
\label{dQ2}
\frac{\partial }{\partial l}\overline{Q^{2}}(E,l)+\Sigma (E)\overline{Q^{2}}(E,l)-\int ^{E}_{0}\Sigma (E_{\gamma };E)\overline{Q^{2}}(E-E_{\gamma },l)dE_{\gamma }=\gamma (E)+2\int ^{E}_{0}E_{\gamma }\Sigma (E_{\gamma };E)\overline{Q}(E-E_{\gamma },l)dE_{\gamma },
\end{equation}
 
\[
\overline{Q^{2}}(E,l=0)=0.\]
 Quantities \( \beta (E) \) and \( \gamma (E) \) in (\ref{sec:Q1}) and (\ref{dQ2})
are the moments of differential cross section \( \Sigma (Q;E): \) 
\[
\beta (E)=\int _{0}^{E}E_{\gamma }\Sigma (E_{\gamma };E)dE_{\gamma },\]
 
\[
\gamma (E)=\int _{0}^{E}E_{\gamma }^{2}\Sigma (E_{\gamma };E)dE_{\gamma }.\]

\section{Continuous slowing down approximation }

\qquad If differential cross section \( \Sigma (E_{\gamma };E) \) is rapidly decreasing
function of the variable \( E_{\gamma } \) the collision integrals in (\ref{dPn}),
(\ref{sec:Q1}), (\ref{dQ2}) can be transformed by the Taylor expansion of integrands:

\[
P(n-1|E-E_{\gamma },l)\approx P(n-1|E,l)-E_{\gamma }\frac{\partial }{\partial E}P(n-1|E,l),\]
 
\[
\overline{Q}(E-E_{\gamma },l)\approx \overline{Q}(E,l)-E_{\gamma }\frac{\partial }{\partial E}\overline{Q}(E,l),\]
 
\[
\overline{Q^{2}}(E-E_{\gamma },l)\approx \overline{Q^{2}}(E,l)-E_{\gamma }\frac{\partial }{\partial E}\overline{Q^{2}}(E,l).\]
 Such transform leads to the equations: 
\begin{equation}
\label{sec:Pn-sl}
\frac{\partial }{\partial l}P(n|E,l)+\Sigma (E)P(n|E,l)=\Sigma (E)P(n-1|E,l)-\beta (E)\frac{\partial }{\partial E}P(n-1|E,l),
\end{equation}
 
\begin{equation}
\label{sec:Q1-sl}
\frac{\partial }{\partial l}\overline{Q}(E,l)+\beta (E)\frac{\partial }{\partial E}\overline{Q}(E,l)=\beta (E),
\end{equation}

\begin{equation}
\label{sec:Q2-sl}
\frac{\partial }{\partial l}\overline{Q^{2}}(E,l)+\beta (E)\frac{\partial }{\partial E}\overline{Q^{2}}(E,l)=\gamma (E)+2\beta (E)\overline{Q}(E,l)-2\gamma (E)\frac{\partial }{\partial E}\overline{Q}(E,l).
\end{equation}

It follows from the Eq.(\ref{sec:Pn-sl}) that in \( \Sigma (E)\approx const \)
approximation \( P(n|E,l) \) is the Poisson distribution:

\begin{equation}
\label{Pn}
P(n|E,l)=\exp (-l\Sigma (E))\frac{(l\Sigma (E))^{n}}{n!}
\end{equation}
with the mean number (multiplicity) of emitted photons equals to

\begin{equation}
\label{n}
\overline{n}(E,l)=\Sigma ^{\infty }_{n=0}nP(n|E,l)=l\Sigma (E).
\end{equation}
To get the solutions of (\ref{sec:Q1-sl}), (\ref{sec:Q2-sl}) quantities \( \beta (E) \)
and \( \gamma (E) \) have to be determined.

\section{Numerical data for bremsstrahlung}

\qquad For a qualitative analysis of the multiphoton bremsstrahlung we used the simplified
relativistic Bethe-Heitler formula \cite{2} for the differential cross section:

\begin{equation}
\label{sigQ}
\Sigma (E_{\gamma };E)=\frac{1}{tE_{\gamma }}(\frac{4}{3}(1-\frac{E_{\gamma }}{E})+\frac{E_{\gamma }^{2}}{E^{2}}),(E^{(c)}_{\gamma }\leq E_{\gamma }\leq E),
\end{equation}
 where \( t \) is the radiation length of the material passed by electron and
\( E_{\gamma }^{(c)}=\alpha E \) is the cut off energy. The cut of for the bremsstrahlung
photon spectra is due to the density effect \cite{5} which suppresses significantly
the soft part of the spectrum. Parameter \( \alpha  \) can be estimated by
the formula

\[
\alpha \sim \frac{\hbar \omega _{p}}{mc^{2}},\]
 where \( \hbar \omega _{p} \) is the plasmon energy of the target material,
\( mc^{2}=511 \) keV is the rest energy of an electron. 

Such \( \Sigma (E_{\gamma };E) \) leads to the energy independent cross section

\begin{equation}
\label{sigma}
\Sigma (E)\approx \frac{1}{t}(\frac{4}{3}\ln \frac{1}{\alpha }-\frac{5}{6})
\end{equation}
 and to the approximate formulas for \( \beta  \) and \( \gamma : \)

\[
\beta (E)\approx \frac{E}{t},\]

\[
\gamma (E)\approx \frac{E^{2}}{2t}.\]

One can see from (\ref{n}), (\ref{sigma}) that the photon multiplicity \( \overline{n} \)
doesn't depend on the electron energy and is determined by the target thickness
only.

Substitution of approximate expression for \( \beta  \) and \( \gamma  \)
into (\ref{sec:Q1-sl}), (\ref{sec:Q2-sl}) and solution of these equations
gives the formulas for the moments:

\begin{equation}
\label{sec:Q1f}
\overline{Q}(E,l)=E(1-\exp (-\frac{l}{t})),
\end{equation}
 
\[
\overline{Q^{2}}(E,l)=\frac{E^{2}}{4}(3-4\exp (-\frac{l}{t})+\exp (-\frac{2l}{t}))\]
 and for the variance:

\begin{equation}
\label{sec:Q2f}
\sigma _{Q}^{2}(E,l)=\overline{Q^{2}}(E,l)-\overline{Q}^{2}(E,l)=\frac{E^{2}}{4}(4\exp (-\frac{l}{t})-3\exp (-\frac{2l}{t})-1).
\end{equation}
 It is seen from (\ref{sec:Q1f}), (\ref{sec:Q2f}), that for \( \frac{l}{t}\ll 1 \)
the moments of bremsstrahlung differential cross section \( \beta  \) and \( \gamma  \)
can be expressed in terms of the moments of the energy loss distribution \( \overline{Q} \)
and \( \sigma _{Q}^{2} \) \( : \)

\[
\overline{Q}(E,l)\approx l\frac{E}{t}=l\beta (E),\]

\[
\sigma _{Q}^{2}\approx l\frac{E^{2}}{2t}=l\gamma (E),\]
which can be determined in experiments where the characteristics of energy loss
distribution are measured.

Formulas (\ref{n}), (\ref{sec:Q1f}) and (\ref{sec:Q2f}) were used to calculate
\( \overline{n,} \) \( \overline{Q} \) and \( \sigma _{Q}^{2} \) for 8 GeV
electron bremsstrahlung in tungsten target. The results were compared with the
data of Monte Carlo simulation using the GEANT code \cite{6}. The Monte Carlo
method was used for calculation of distribution of electrons over the number
of emitted photons. The calculations show that the distribution strictly follows
to the Poisson law (\ref{Pn}). Mean number of photons (Fig.1) agrees with (\ref{n}),
(\ref{sigma}) if \( \alpha \approx 1.1\cdot 10^{-5} \). This value of \( \alpha  \)
corresponds to the cut of energy \( E^{(c)}_{\gamma }\approx 88keV. \) Fig.2
shows that the energy distribution \( P(E_{\gamma };E,l) \) of emitted photons
(the number of photons per unit interval of \( E_{\gamma } \)) really vanishes
for \( E_{\gamma }<E^{(c)}_{\gamma }. \)

\begin{figure}[ht]
\parbox[t]{0.45\textwidth}{
\includegraphics[trim= 1.5cm 6cm 2cm 6.5cm,clip,width=8.5cm,height=8.5cm]{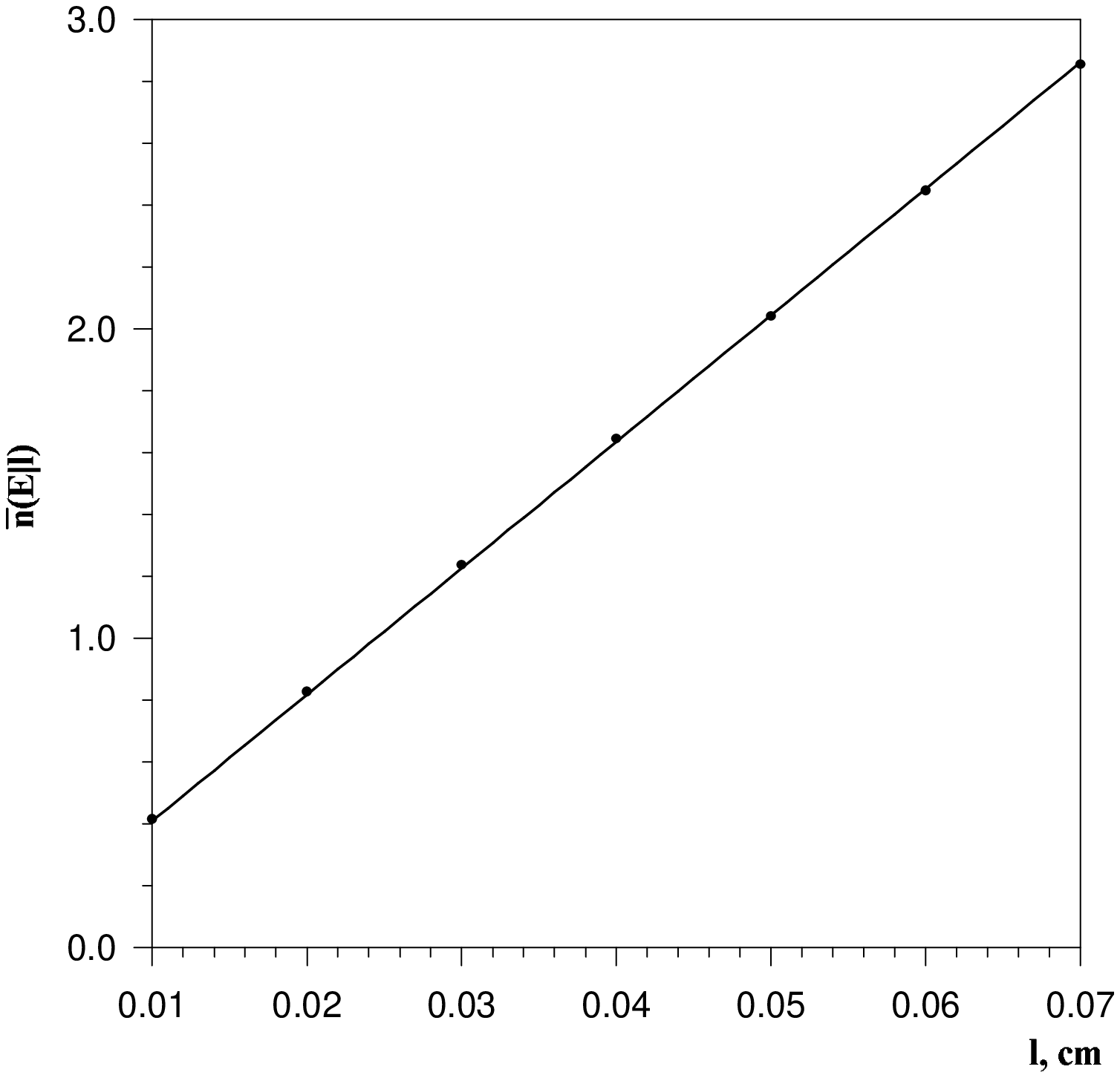}
\caption{Mean number of photons. \break
solid line --- Eq. (10), points --- simulation.}}
\parbox[t]{0.45\textwidth}{
\includegraphics[width=8.5cm,height=8.5cm]{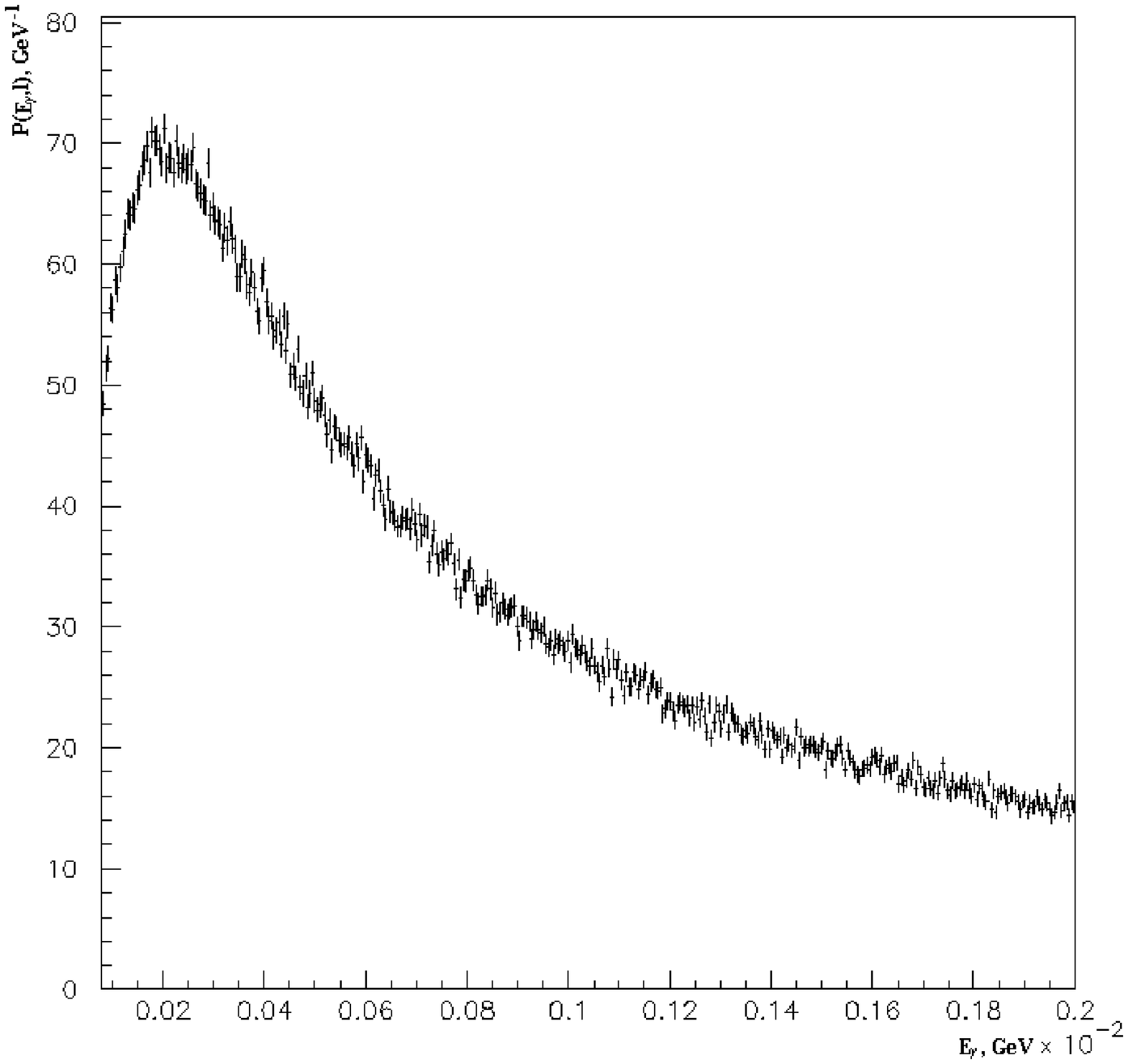}
\caption{Energy distributions of emitted photons at l=0.01cm}}
\end{figure}

For small \( l \) when multiphoton effect can be neglected the energy distribution 
 \( P(E_{\gamma };E,l) \) can be
expressed in terms of the differential cross section \( \Sigma (E_{\gamma };E): \)
\( P(E_{\gamma };E,l)=l\Sigma (E_{\gamma };E). \) Fig.3 shows that the distribution
\( P(E_{\gamma };E,l) \) obtained by simulation agrees with that one calculated
with using the formula (\ref{sigQ}).

Data on \( \overline{Q}(E,l) \) are given in Fig.4 as well as the mean energy
of emitted photons: \( \overline{E_{\gamma }}=\frac{\overline{Q}}{\overline{n}} \).
It is seen that \( \overline{E_{\gamma }} \) slowly decreases with \( l \)
because of electron energy decreasing. The variance of energy loss \( \sigma _{Q}^{2}(E,l) \)
is shown in Fig.5. One can see that the analytical formulas (\ref{sec:Q1f}), (\ref{sec:Q2f})
for \( \overline{Q} \) and \( \sigma _{Q}^{2} \) agrees with the data of Monte
Carlo simulation quite good.

The Monte Carlo code was used to calculate the probability density function
\( P(Q|E,l) \) describing the distribution of electrons over energy loss \( Q \)
and the ratio of \( P(Q|E,l) \) to the energy spectra. The results are given
in Fig.6 which displays the difference between radiation loss spectrum from 
finite thickness target and intensity spectrum of bremsstrahlung in the field of
single nucleus.
It is clear that multiphoton emission depresses the low energy part
of the distribution and increases the high energy part. The effect rises
with the target thickness increasing.

This work is supported by the Grant of the Ministry of General and Professional
Education of the Russian Federation.
\newpage
\begin{figure}[ht]
\parbox[t]{0.46\textwidth}{
\includegraphics[width=8.5cm,height=8.5cm]{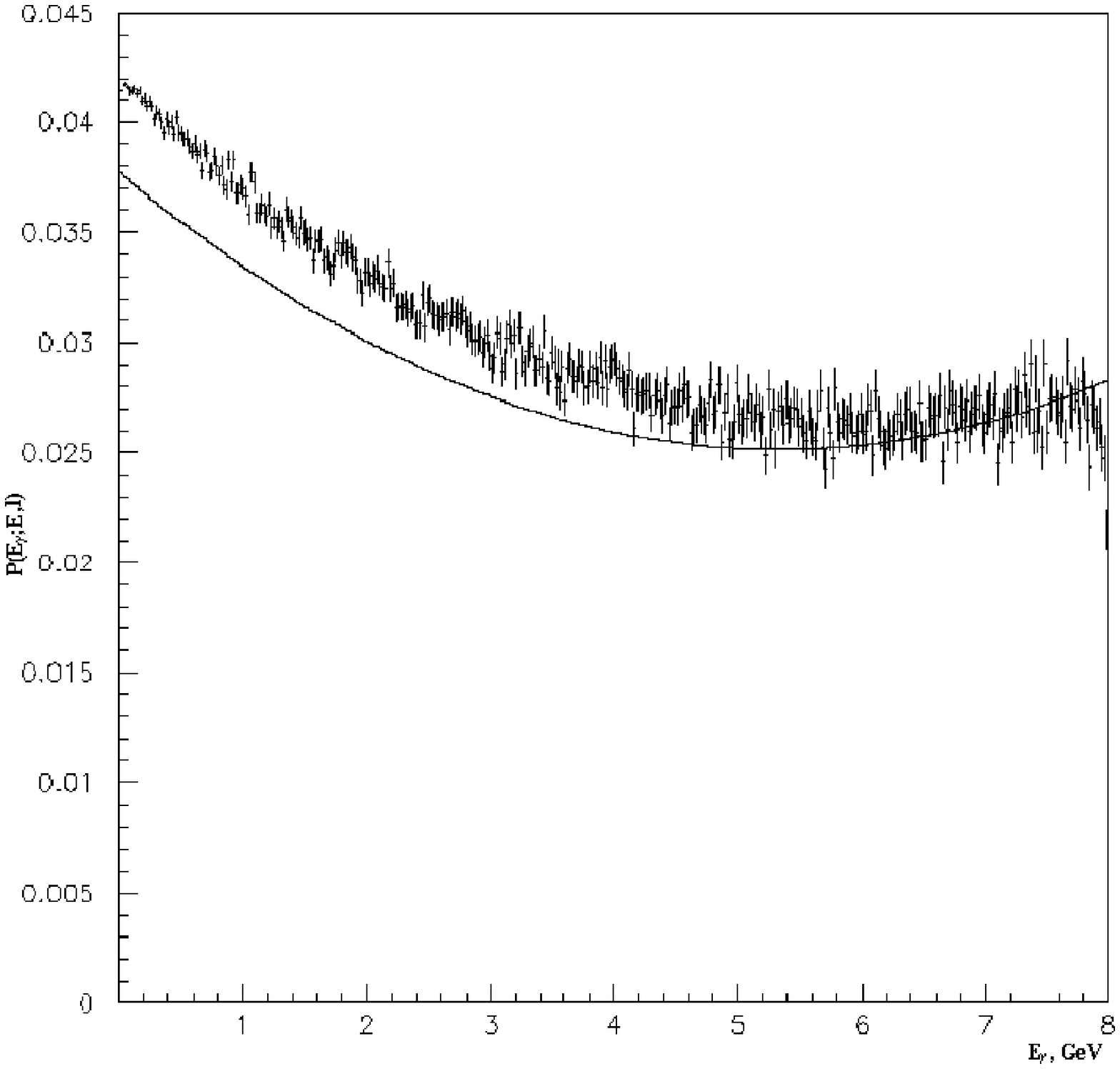}
\caption{Comparison of photons energy distribution
corresponding to Eq.\ (11) (solid line) with simulation at $l=0.01cm$}}
\hspace{3mm}
\parbox[t]{0.4\textwidth}{
\includegraphics[trim= 1.7cm 6cm 2cm 6cm,clip,width=8.5cm,height=8.5cm]{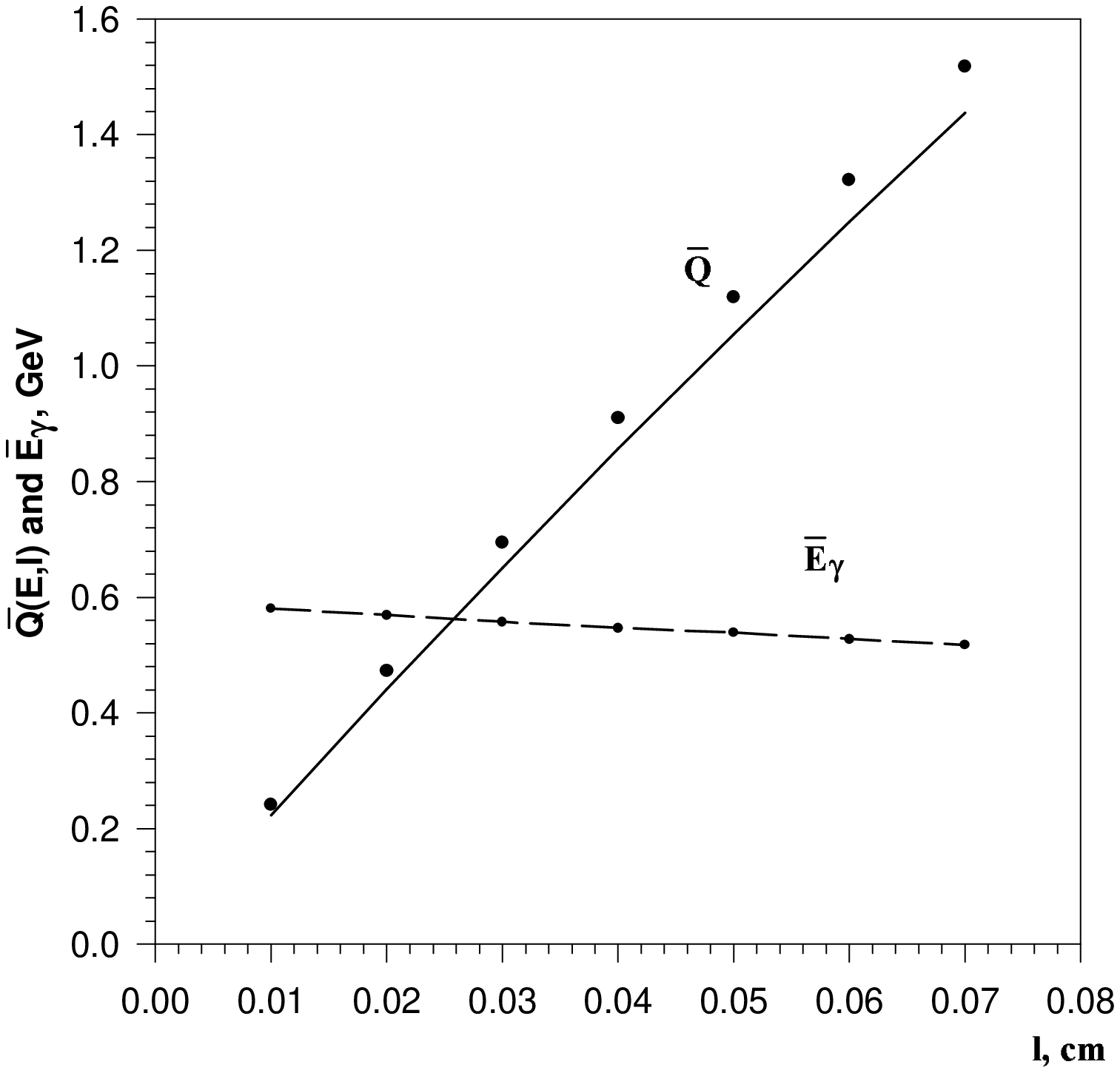}
\caption{Mean energy loss $\overline{Q}(E,l)$ and 
mean energy of emitted
photons $\overline{E_\gamma}$. \break
Solid line --- Eq.\  (13), points --- simulation}}
\begin{center}
\includegraphics[trim= 2cm 6cm 2cm 6cm,clip,width=8.5cm,height=8.5cm]{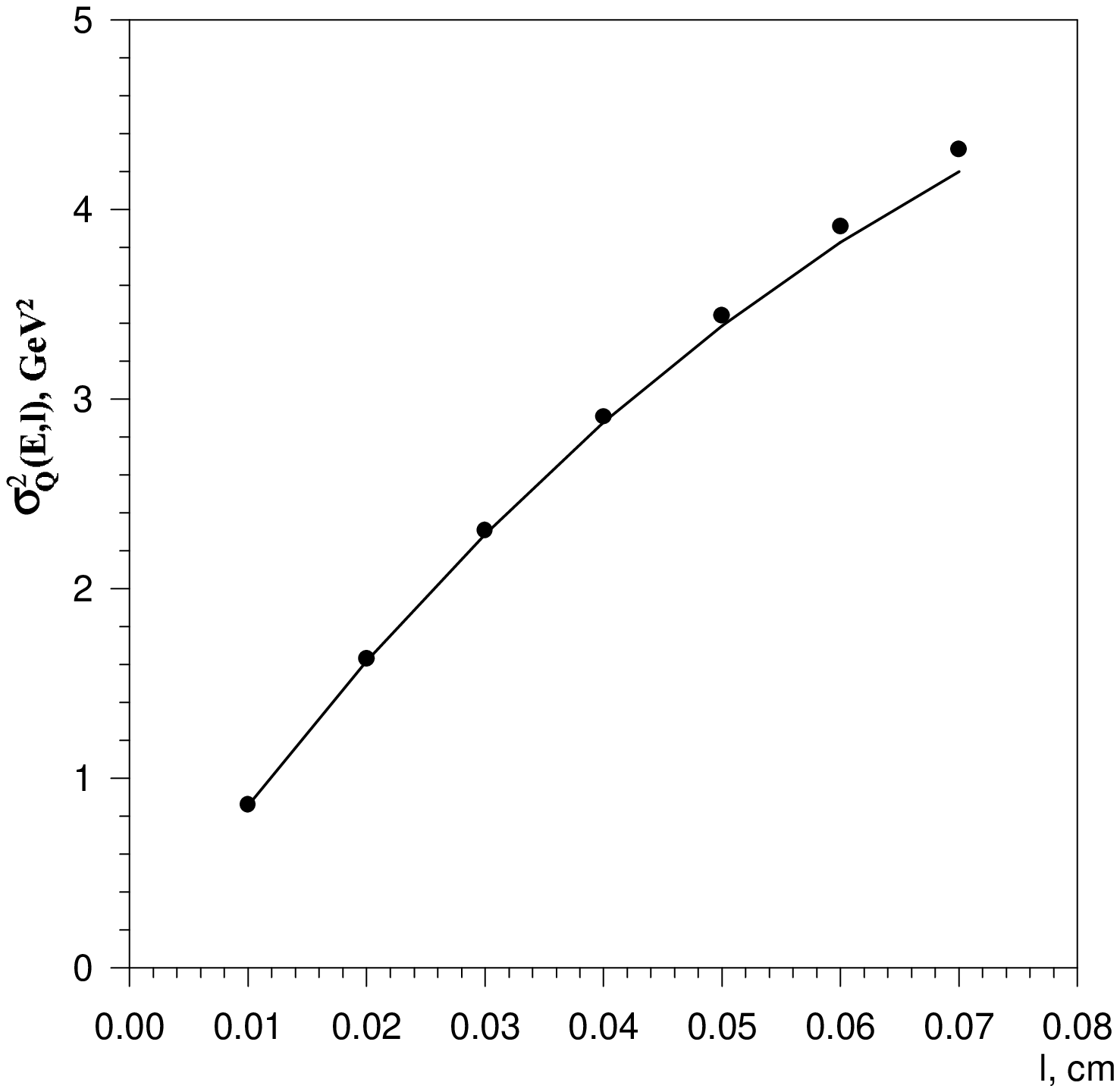}
\end{center}
\caption{Variance of energy loss$.$ 
Solid line --- Eq.\  (14), points --- simulation}
\end{figure}
\begin{figure}
\includegraphics[width=18cm,height=18cm]{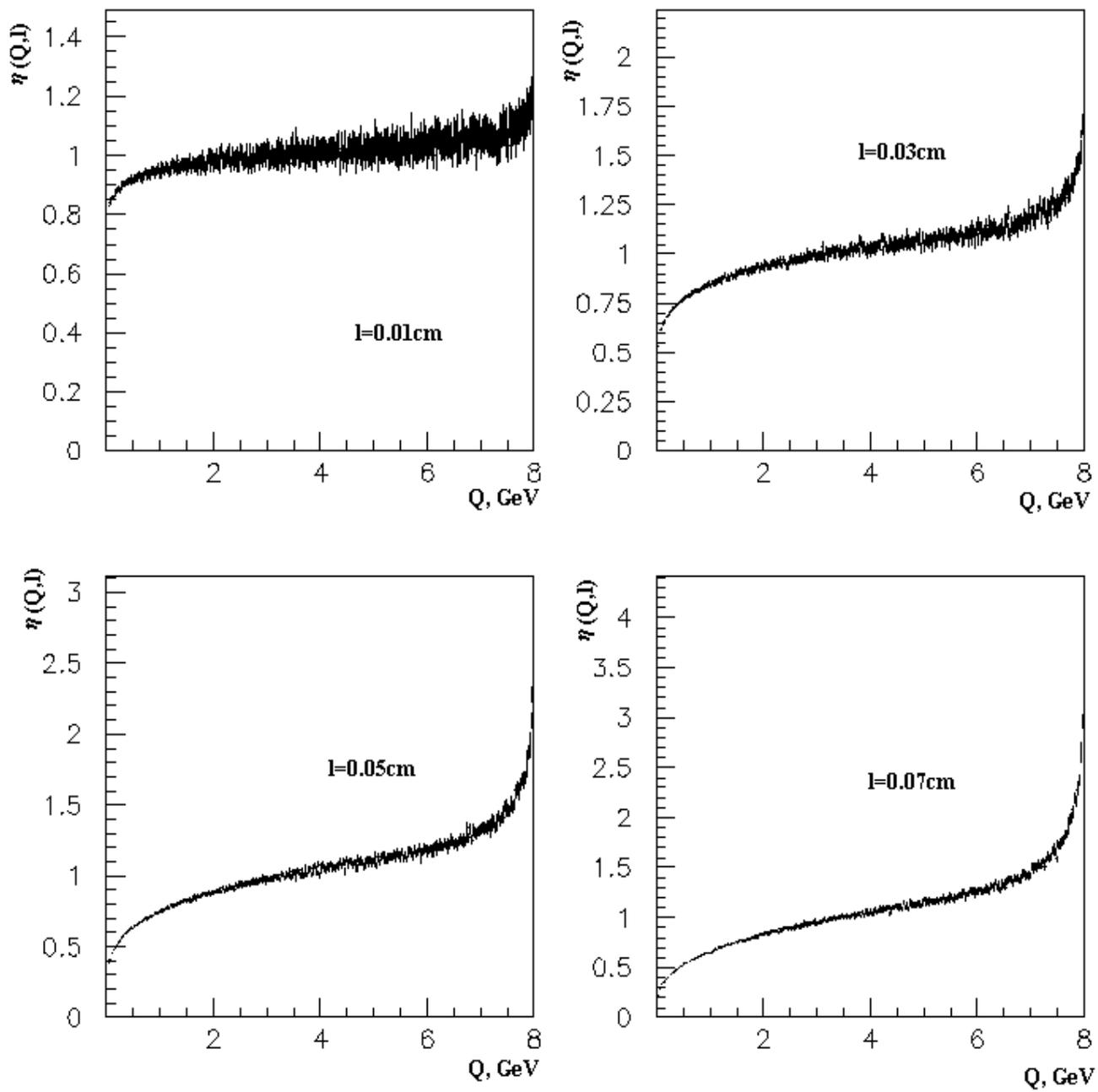}
\caption{The ratio $\eta(Q,l)=\frac{P(Q|E,l)}{P(E_\gamma=Q|E,l)}$}
\end{figure}

\end{document}